\documentclass[aps,prl,twocolumn,showpacs,preprintnumbers,amsmath,amssymb,superscriptaddress]{revtex4}
\usepackage{color}
\usepackage{graphicx}
\usepackage{dcolumn}
\usepackage{bm}
\usepackage[hypertex]{hyperref}
\newcommand{\nc}{\newcommand}
\nc{\be}{\begin{eqnarray}}
\nc{\ee}{\end{eqnarray}}
\nc{\bea}{\begin{eqnarray}}
\nc{\eea}{\end{eqnarray}}
\nc{\bean}{\begin{eqnarray*}}
\nc{\eean}{\end{eqnarray*}}
\nc{\mb}{\mbox}
\nc{\rnc}{\renewcommand}
\nc{\vk}{\mb{\bf k}}
\nc{\vp}{\mb{\boldmath$p$}}
\nc{\rr}{\mb{\boldmath$r$}}
\nc{\RR}{\mb{\boldmath$R$}}
\nc{\vz}{\hat {\mb{\bf z}}}
\nc{\vj}{\mb{\boldmath$j$}}
\nc{\vg}{\mb{\boldmath$g$}}
\nc{\x}{\mb{\boldmath$x$}}
\nc{\A}{\mb{\boldmath$A$}}
\nc{\va}{\mb{\boldmath$a$}}
\nc{\vq}{\mb{\boldmath$q$}}
\nc{\vn}{\mb{\boldmath$n$}}
\nc{\vs}{\mb{\boldmath$\sigma$}}
\nc{\vt}{\mb{\boldmath$\tau$}}
\nc{\vpi}{\mb{\boldmath$\pi$}}
\nc{\nab}{\bm{\nabla}}
\nc{\X}{\sf x}

\begin{document}


\title{
Field-induced Kosterlitz-Thouless transition
in the $N=0$ Landau level of graphene
}

\author{Kentaro Nomura}
\affiliation{
Department of Physics, Tohoku University, Sendai, 980-8578, Japan
            }
\author{Shinsei Ryu}
\affiliation{
  Department of Physics, University of California at Berkeley,
Berkeley, CA 94720, USA
            }
\author{Dung-Hai Lee}
\affiliation{
Department of Physics, University of California at Berkeley,
Berkeley, CA 94720, USA
}
\affiliation{
Material Science Division, Lawrence Berkeley National Laboratory,
Berkeley, CA 94720, USA}

\date{\today}

\begin{abstract}
At the charge neutral point, graphene exhibits a very unusual high resistance metallic 
state and a transition to a complete insulating phase in a strong magnetic field.
We propose that the current carriers in this state are the charged vortices of
the XY valley-pseudospin order-parameter, 
a situation which is dual to a conventional thin superconducting film.
We study energetics and
the stability of this phase in the presence of disorder.
\end{abstract}

\maketitle

\noindent

The initial experiments of the quantum Hall effect (QHE) in monolayer graphene
discovered the quantum Hall plateaus
$
\sigma_{xy} =
4
\left(N +\frac{1}{2}
\right)
(e^2/h)
$
($N \in \mathbb{Z}$)
at filling factors
$\nu=2\pi\ell_B^2\rho
=
\pm2,\pm6,\pm10, \cdots$\cite{Review}.
Here,
$\ell_B$ is the magnetic length, 
$\rho$ the carrier density measured from the charge neutral point,
and
the factor of $4$ arises from the spin and valley ($K$ and $K'$) degeneracy.
Recent experiments under stronger magnetic fields, on the other hand, 
showed additional plateaus at $\nu=0,\pm1,\pm4$
\cite{Zhang_2006,Jiang_2007}.
Further experiments\cite{Ong08PRL}
at $\nu=0$ in high quality samples
revealed a rapid divergence of the longitudinal
resistance $R_{xx}$
at a critical field $B_c$. 
Interestingly such divergence fits the Kosterlitz-Thouless (KT)\cite{KT_review} form
$R_{xx}\sim e^{a/\sqrt{B_c-B}}$
over three decades of the resistance\cite{Ong08PRL}.
Moreover,  for $B<B_c$ the resistance saturates at low
temperature to a value much larger than the quantum of resistance
\cite{Ong08PRL},
a behavior qualitatively different 
from conventional thermally activated 
transport in strong magnetic fields.

Since the critical field $B_c$ lies in the regime where the 4-fold degeneracy of the $N=0$ Landau levels (LLs)
is split, it is important to understand the cause of such splitting.
There are several theoretical proposals.
The common theme is the observation that the exchange effect of the long-range part of the Coulomb interaction\cite{QHF_review} favors the spontaneous polarization of the real and/or valley-spins
\cite{Nomura_2006,Alicea_2006,Goerbig_2006,Sheng_2007, CPE2006, CPE2007,
Gusynin_2006,Herbut_2006,Fuchs_2006,Ando_1994,Hatsugai_2008,Khveshchenko_2001}.
(the valley-spin is a SU(2) pseudospin variable,
which we denote by $\boldsymbol{T}$; 
its $z$-component $T_z=+1$ ($-1$)
corresponds to $K$ ($K^{\prime}$), respectively.)
In one of the proposals, it is argued
that the Zeeman energy favors the polarization of the real instead of the valley-spin\cite{CPE2006,CPE2007}. 
In another
it is argued that the short-range part of the Coulomb interaction favors  the spontaneous polarization of the valley-spin so that $K$ and $K^\prime$ becomes unequally populated\cite{Khveshchenko_2001,Alicea_2006,Gusynin_2006,Herbut_2006,Fuchs_2006, Sheng_2007}. 
Because the $N=0$ Landau states associated with $K'$ and $K$ localize
on complementary sublattices ($A$ and $B$, respectively)
this amounts to a charge-density-wave (CDW) modulation which breaks the $A$-$B$ sublattice symmetry.

\begin{figure}[b]
\begin{center}
\includegraphics[width=0.48\textwidth]{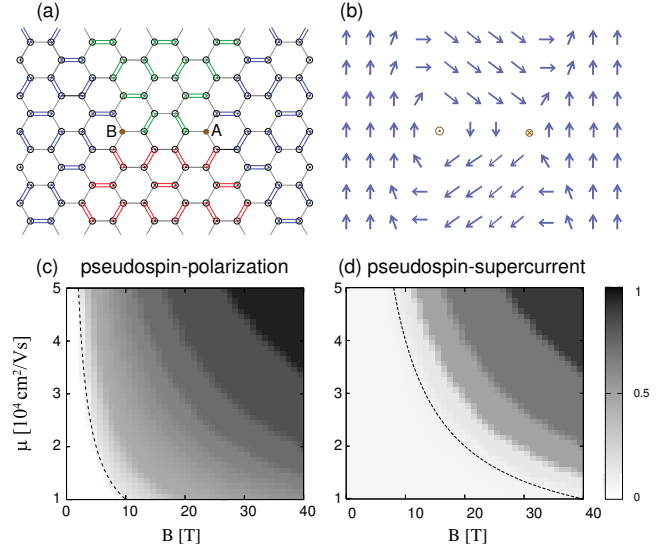}
\caption{
(Color online)
(a) The Kekule bond-density wave order with two defects 
marked by a filled circle. The defects are charged as they support 
a midgap electron state. 
(b) The U(1) phase $\phi=\tan^{-1}(T_y/T_x)$ corresponding to the bond order pattern (a). 
(c) The valley-spin polarization ratio as a function of magnetic field $B$ and sample mobility $\mu$.
(d) Same as (a) but for the pseudospin-supercurrent $j^z_{\rm sc}/j^{z{\rm (clean)}}_{\rm sc}$, where $j^{z {\rm (clean)}}_{\rm sc}$
is the value in the clean limit.
The number of orbitals per valley and per spin is $N_{\phi}=50$.
}
\label{HF figure}
\end{center}
\end{figure}

The type of diverging resistance observed in Ref.\ [\onlinecite{Ong08PRL}] is
difficult to account for in the real spin polarization scenario
since in this scenario
there are spin-filtered counter-propagating edge states 
that give rise to a metallic conductance of $2e^2/h$ at $\nu=0$
\cite{CPE2006, CPE2007,Shimshoni_2008}.
Similarly in the CDW scenario \cite{Gusynin_2006,Herbut_2006,Fuchs_2006,Khveshchenko_2001} it is difficult to explain 
the KT type resistance divergence,
and the high-resistance metallic state below the critical field
\cite{Ong08PRL}.
In addition to the above, there is a work
(Ref. \cite{Luk_2008}) which claims an explicit valley symmetry breaking term is consistent with the lattice point group symmetry. Like others, this work can not account for the KT behavior.

Motivated by the KT behavior
and the highly resistive metallic state,\
we propose an alternative scenario:
the degeneracy splitting at $\nu=0$ is due to
a spontaneous ordering of the pseudospin
on the $T_x-T_y$ plane (XY pseudospin ferromagnet).
This involves a spontaneously
generated hybridization between the $N=0$ LLs associated with
$B\ (K)$ and $A\ (K^\prime)$,
and is represented by the ground state
wavefunction
\bea
|\Psi\rangle =\prod_{m,s=\uparrow\downarrow}\frac{1}{\sqrt{2}}
\left[
c^{\dag}_{Kms}+e^{i\phi}c^{\dag}_{K'ms}
\right]|0\rangle,
\label{gs wfn}
\eea
where $c^{\dag}_{\tau ms}$
is the creation operator
for an electron in $m$-th $N=0$ LL orbital
at valley $\tau = K ,K'$ with real spin $s=\uparrow,\downarrow$.
This type of order also breaks the lattice translation symmetry
due to the mixing of $B$ and $A$,
and represents a bond-density-wave of some kind (Kekule order)
[see Fig.\ \ref{HF figure}(a)]
\cite{Mintmire_1992,Chamon_2000, Hou_2007,Ando_1994,Hatsugai_2008}.
The phase $\phi$ of this hybridization matrix element is
the $\textrm{U}(1)$ phase angle 
representing the direction in $T_x-T_y$ plane,
$\boldsymbol{T}=(\cos\phi,\sin\phi,0)$,
and associated with the sliding degrees of freedom of this density-wave.
The low-energy charged excitations are vortices and antivortices 
[Fig.\ \ref{HF figure}(b)].
We study their binding-unbinding transition driven by magnetic fields or disorder
(see below).

We now take a look at,
piece by piece,
the Hamiltonian for graphene in a strong magnetic field
and the associated energy scales,
to address the plausibility of the inter-valley coherent state.
(1) At the charge neutral point the LL
separation is
$
\sqrt{2}\hbar v_F^{}/\ell_B
\simeq 400\sqrt{B[\mathrm{T}]}[\mathrm{K}]$,
which is the largest energy scale of the problem. Therefore in the rest of the paper we perform projection onto
the $N=0$ LLs.

(2) The Coulomb interaction $H_C$,
which is the second largest energy scale of the problem,
is approximately symmetric under rotation in the combined space of real and valley-spins.
The exchange energy 
is
$
E^{\mathrm{ex}}_C
\sim \sqrt{\pi/2}(e^2/\epsilon \ell_B)\simeq 120\sqrt{B[\mathrm{T}]}[\mathrm{K}]$
\cite{QHF_review}.

As a result it favors the polarization of the SU(4) spin albeit it does not care whether the polarization should occur in the real spin, or valley-spin, or some combination of both\cite{Nomura_2006}.

(3) We now describe the parts of the Hamiltonian
which break the SU(4) symmetry,
$H_{\mathrm{SB}}=\int d^2 r\, \mathcal{H}_{\mathrm{SB}}$,
\be
\mathcal{H}_{\mathrm{SB}}
\!\!&=&\!\!
-\frac{1}{2}\Delta_z S_z
-U_0 |\boldsymbol{S}|^2-U_z T_z^2-U_{\perp} (T_x^2+T_y^2). 
\label{HSB}
\ee
Here $\boldsymbol{S}$ is the real spin operator. 
The first term in $H_{\mathrm{SB}}$
represents the Zeeman energy where
$
\Delta_z\equiv g\mu_B
B
\simeq
1.3\times(B[\mathrm{T}])[\mathrm{K}].
$
The short-range part of the Coulomb interaction is not
SU(4) symmetric and gives rise to $U_0$ and $U_z$.
They can be estimated from the on-site and the nearest neighbor interactions,
and are smaller than $E^{\mathrm{ex}}_C$ 
by a factor $a/ \ell_B$\cite{Alicea_2006,Goerbig_2006},
where $a$ is the lattice constant.
While $U_0$ favors the real spin polarization, $U_z$ favors the
CDW phase $(T_z\neq 0)$\cite{Alicea_2006,Goerbig_2006,Sheng_2007}.

\begin{table}[bp]
\begin{ruledtabular}
\begin{tabular}{c|l|cc}
& residual symmetry  &   energy scale & \\
& [(spin) $\times$ (valley)] &   & \\   \hline
$\Delta_z$ & ${\rm No} \times \mathrm{SU}(2)$
                             & $ 1.3 [\mathrm{K}]\times B[\mathrm{T}]$\ \  \cite{Jiang_2007} & \\
$U_0$ & $\mathrm{SU}(2)\times \mathrm{SU}(2)$ &  $ 1.0 [\mathrm{K}]\times B[\mathrm{T}]$\ \  \cite{Alicea_2006}  & \\
$U_z$ & $\mathrm{SU}(2)\times \mathbb{Z}_2$ (CDW) &
  $0.5 [\mathrm{K}]\times B[\mathrm{T}]$\ \ \cite{Alicea_2006}  & \\
$U_{\perp}$ & $\mathrm{SU}(2)\times \mathrm{U}(1)$ (Kekule)
                              & $2.0 [\mathrm{K}]\times B[\mathrm{T}]$\  \cite{Ando_1994} & \\
        \end{tabular}
\end{ruledtabular}
\caption{
\label{symmetry breaking terms}
SU(4) symmetry breaking terms,
with the pattern of symmetry breaking and the energy scales.
}
\end{table}

On the other hand, $U_{\perp}$ term can arise from
the electron-phonon interaction.
One such example is the in-plane optical mode at $K$ point,
whose interaction with electrons
can schematically
be represented as\cite{Ando_1994,Sasaki_2008}
\begin{eqnarray}
H_{\perp} =f\int d^2r\, \,
\boldsymbol{u}
\cdot \left(
\psi^{\dag} \vt\sigma_x \psi
\right)
+
\frac{N_C k}{2} {\mb{\boldmath$u$}}^2,
\end{eqnarray}
where $\boldsymbol{u}=(u_x,u_y)$ represents
the (uniform) Kekule-type distortion of the lattice,
$k$ measures the elastic energy,
and $N_C$ is the total number of carbon atoms.
The two sets of Pauli matrices,
$\{\sigma_{x,y,z,0}\}$ and $\{\tau_{x,y,z,0}\}$,
act on sublattice ($A,B$) and valley ($K,K'$),
respectively;
$\psi^{\dag}(\tau_{x}\pm i\tau_y)\sigma_x \psi \propto T_x\pm iT_y\propto e^{\pm i\phi}$
serves as the $\mathrm{U}(1)$ order-parameter of the Kekule bond density wave. 
Upon integrating out the phonon, this give rise to the $U_{\perp}$ term with
$
U_{\perp} \sim
2.0 \times (B[\mathrm{T}]) [\mathrm{K}].
$
 Note that
$U_{\perp}$ is comparable to $\Delta_z$. This can be traced back to the strong coupling between the $K$ phonon and electrons \cite{kekule2000,phonon2007}.
Out-of-plane lattice distortion is studied in Ref.\ [\onlinecite{Fuchs_2006}],
and shown to contribute to $U_z$
which
are much weaker than that associated with in-plane modes in graphene\cite{Sasaki_2008}. 
The SU(4) breaking terms are summarized in Table \ref{symmetry breaking terms}.

Since the SU(4) symmetric part of the Coulomb interaction
is much stronger than
the symmetry breaking parts $H_{\mathrm{SB}}$
it is the former that sets the basic energy scale for
the SU(4) symmetry breaking.
The symmetry breaking terms simply
select
the way the SU(4) symmetry is broken:
they determine the nature of the ordered phase.
Although $U_0,U_z,U_{\perp}$ 
in Table \ref{symmetry breaking terms}
all have similar energy scales, 
it suggests
the $\mathrm{U}(1)$ broken inter-valley coherent state (\ref{gs wfn})
is a reasonable
candidate for lifting the degeneracy of the $N=0$ LLs.

We now describe the field-induced transition at zero-temperature
using the self-consistent Hartree-Fock (HF) theory.
To account for competition between
interaction
 and disorder effects,
we allow the XY pseudospin order-parameter to be
spatially inhomogeneous.
The matrix element of the HF Hamiltonian, in the Landau gauge, 
can be written in the form\cite{Sinova_2000}, 
\begin{eqnarray}
\langle m\sigma|H_{\mathrm{HF}}|m'\sigma'\rangle
=
\sum_{\boldsymbol{q}}
e^{iq_x X_{m'}}\delta_{q_y\ell^2_B,X^{\ }_{m}-X_{m'}}
\qquad \quad
\nonumber \\
\times
\Big[
U_{\sigma\sigma'}^H({\vq})
+U_{\sigma\sigma'}^F({\vq})
+U^{XY}_{\sigma\sigma'}({\vq})
+\delta_{\sigma\sigma'}U_{\rm imp}({\vq})
\Big].
\label{HFhamiltonian}
\end{eqnarray}
Here the system size is $L\times L=2\pi\ell_B^2N_{\phi}$, 
$X_m= 2\pi \ell_B^2m/L$ $(m=1,2,\cdots,N_{\phi})$,
and $\sigma=1,\ldots, 4$ is the index for (real)spin and valley.
The Hartree and Fock potentials,
and the anisotropic interaction are given as follows:
\bea
U_{\sigma\sigma'}^{H}(\vq)
\!\!&=&\!\!
\frac{\delta_{\sigma\sigma'}}{2\pi\ell_B^2}V_C(\vq)e^{-{q^2\ell_B^2\over 2}}
\sum_{\sigma''}\Delta_{\sigma\sigma''}(\vq),
\label{eq: HF hamiltonian 2}
\\
U_{\sigma\sigma'}^{F}(\vq)
\!\!&=&\!\!
-\frac{1}{L^2} \sum_{\boldsymbol{p}}
V_C(\vp)
e^{-{p^2\ell_B^2\over 2}
+i\boldsymbol{q}\times \boldsymbol{p}\ell_B^2}
\Delta_{\sigma\sigma'}(\vq),
\nonumber \\
U_{\sigma\sigma'}^{XY}(\vq)
\!\!&=&\!\!
\frac{-U_{\perp}}{2\pi}
\sum_{\sigma_1\sigma_2}
\Delta_{\sigma_1\sigma_2}(\vq) e^{-{q^2\ell_B^2\over 2}}
\sum_{j=x,y}
{\hat T}^j_{\sigma\sigma'}{\hat T}^j_{\sigma_1\sigma_2},
\nonumber
\eea
where
$V_C(\vq)=2\pi e^2/\epsilon q$ and 
${\hat T}^{j}= \sigma_0 \otimes\tau_{j}$.
The SU(4) order-parameter in Eqs.\ (\ref{HFhamiltonian},\ \ref{eq: HF hamiltonian 2})
is determined self-consistently from
$
\Delta_{\sigma'\sigma}({\vq})=
N_{\phi}^{-1}\sum_{mm'}\langle c_{m\sigma}^{\dag}c_{m'\sigma'}^{} \rangle e^{-iq_xX_m}\delta_{q_y\ell^2_B,X_{m'}-X_m}.
$
The disorder potential $U_{\rm imp}(\vq)$
is given in terms of 
charged impurities located randomly at ${\bf R}_I$
by
$
U_{\rm imp}(\vq)=
L^{-2}
\sum_{I=1}^{N_{\rm imp}}e^{i \boldsymbol{q} \cdot {\bf R}_I}V_C(\vq)
e^{-[q^2/4-iq_xq_y/2]\ell^2_B}
$.
The disorder strength is described
by the impurity filling $\nu_{\rm imp}=(h/eB)n_{\rm imp}$, where $n_{\rm imp}=N_{\rm imp}/L^2$ is the impurity density.
In terms of the zero-field mobility of graphene, 
$\mu=\sigma_{xx}/ne\simeq 20 e/hn_{\rm imp}$\cite{Nomura_2006}, 
the disorder strength is determined by a product $B\times\mu$.

In the clean limit, $N_{\rm imp}=0$, 
the ground state is fully pseudospin polarized due to the Coulomb exchange.
In a weak field or in a dirty sample, on the other hand,
the Coulomb interaction plays a minor role
and the four (nearly) degenerated LLs are equally occupied.
As shown in Fig.\ \ref{HF figure} (c),
the pseudospin polarization ratio diminishes in the weak field and low mobility limit. In particular
when $B<B^*\simeq 10/(\mu [10^4{\rm cm^2/Vs}])[{\rm T}]$ the pseudospin symmetry is restored by disorder.
 Here $B^*$, referred to the dashed line in Fig.\ \ref{HF figure} (c), plays the role of the mean-field critical field.

\begin{figure}[bbp]
 \begin{center}
 \includegraphics[scale=0.26]{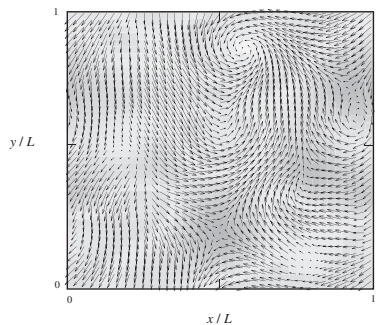}
 \caption{
A typical local XY pseudospin configuration $(T_x, T_y)$
represented by arrows. The $z$-component is represented by a grey plot with black as $+1$ and white as $-1$.
The number of orbitals per valley and per spin is $N_{\phi}=50$, and the number of impurities is $N_{\rm imp}=20$.
\label{texture}}
 \end{center}
 \end{figure}

Even $B>B^*$, vortices and anti-vortices tend to destroy the stiffness of the XY pseudospin order.
An inkling of this KT transition
could be seen
in the unrestricted HF calculation
as follows.
In the pseudospin XY
(quasi-long-range) ordered phase,
 angle-twisted states which have $\phi=QX_m$ in Eq.\ (\ref{gs wfn}), namely
{\it pseudospin-supercurrent-flowing states},
are metastable, because of the finite stiffness.
Such states can be selected
by artificially starting with an initial order-parameter
$\Delta_{KK'}(\vq)=\delta_{
\boldsymbol{q},Q{\hat {\bf x}}}$
in the initial step of the self-consistency loop.
One can then monitor whether or not
the pseudospin-supercurrent generated can survive
as one iterates the HF calculation\cite{Sinova_unpublished}. 
Charged impurities generate quenched vortex-antivortex pairs
that randomize $\phi$, and hence
the pseudospin-supercurrent is expected to vanish in a weak field and low $\mu$s
(see Fig.\ \ref{texture}.)
In Fig.\ \ref{HF figure} (d),
the pseudospin-supercurrent of the metastable state,
given by \cite{Abolfath_2003}
$
\vj^{z}_{\rm sc}=\frac{-e}{L^2}\sum_{\boldsymbol{q},ss'\tau}\tau[i\vq\times\vz]
V_C(\vq)
\rho_{\tau s}(-\vq)\rho_{-\tau s'}(\vq),
$
is plotted,
where $\tau=\pm 1$ for $K$ and $K'$, and $\rho_{\tau s}(\vq)$ is the density operator for valley $\tau$ and spin $s$.
Here note that pseudospin-supercurrents are charge neutral objects and
thus are not related to true charge-currents. 
We calculate $\vj^{z}_{\rm sc}$ to discriminate whether vortices and antivortices are bound or unbound.
As Fig.\ \ref{HF figure} (d)
indicates, the pseudospin supercurrent drops around
$B_c\simeq  40/(\mu[10^4\ {\rm cm}^2/{\rm Vs}])
\,   [{\rm T}]> B^*$.

When $B\gg B_c$ and when the temperatures is sufficiently lower than the Coulomb exchange energy but still finite, the U(1) phase fluctuations are described by the following classical action\cite{KT_review,QHF_review}
\begin{eqnarray}
S_{\mathrm{XY}}
\!\!&=&\!\!
\frac{\rho_s}{2}  \int d^2x \; ({\bm{\nabla}}\phi)^2.
\label{action_xy}
\end{eqnarray}
The unbinding of the vortex-antivortex pairs triggers the KT transition
from
the pseudo spin XY quasi-long-range ordered
phase to the disorder phase at $B_c$.

Vortices and anti-vortices are charged
\cite{QHF_review, Hou_2007}
and they can contribute to electrical transport. 
The reason why they carry a charge can be understood 
on the honeycomb lattice as follows [Fig.\ \ref{HF figure}(a)]: 
A defect in the Kekule order can be visualized as a $A$ or $B$ 
sublattice site that is not dimerized with neighbors,
and hence supports a midgap state (zero mode).
The presence (absence) of an electron on such site
makes the Kekule vortex or antivortex positively (negatively) charged.
The pseudo spin is pointing $T_z=+1$ ($-1$) at the vortex core
while $T_z=-1$ ($+1$) at the antivortex core.
Hence, the charge and currents generated by vortex excitations are given by
$j_{\mu}=(T_z/\pi)\epsilon_{\mu\nu\lambda}\partial_{\nu}\partial_{\lambda}\phi$,
where $\mu=0,x,y$.

In the XY ordered phase where the vortex and antivortex are bound, 
it is energetically favorable for the $T_z$ of the pair to point in the same direction.
As a result, the bound vortex pairs are charge neutral.
On the other hand, a charged vortex-antivortex pair can be induced and pinned by charged impurities. 
As increasing the number of vortex-antivortex pairs, by increasing the impurity density 
or by decreasing magnetic fields, 
the binding interactions between vortex and antivortex are screened, causing the KT transition.

In the KT disordered phase ($B^* < B < B_{c}$),
the vortices are unbound,
and their diffusion gives rise to
a conductivity given by
$\sigma \propto n_{\mathrm{vtx}} \mu_{\mathrm{vtx}}$
where $n_{\mathrm{vtx}}$ is the density of vortices
and $\mu_{\mathrm{vtx}}$ the mobility.
This vortex conducting mechanism is a two-dimensional analog 
of the soliton conduction mediated by charged defects (domain walls) in polyacetylene\cite{Su-Schrieffer}. 
In the KT disordered phase
$n_{\mathrm{vtx}}\sim 1/\xi^2$
where $\xi$ is the KT correlation length\cite{KT_review}.
Since
$\xi\propto
e^{a/\sqrt{B_c-B}}
$, this gives rise to the KT-divergent resistivity.
This argument closely follows the one used by Halperin and Nelson 
in analyzing the behavior of the electrical conductivity of a thin film 
superconductor above its KT transition\cite{HN}. 
Indeed, our situation is dual to theirs. 
In Ref.\ \cite{HN} the Cooper pair (charge) current exerts 
the magnus force on the vortices and, through the finite vortex mobility, 
induces a vortex current perpendicular to it.  
Since vortex current causes an transverse 
electric field (hence a voltage drop) through the Josephson relation, 
this gives rise to a finite electrical resistivity. 
In our case the vortex is charged, and it is the external electric field that 
induced the vortex (charge) current. 
Thus our electric field plays the role of charge (Cooper pair) current 
in Ref.\ \cite{HN}, 
while our charge (vortex) current plays the role of electric field 
in Ref.\ \cite{HN}.
As the result, electrical conductivity in Ref.\ \cite{HN} should be 
translated into electric resistivity $\rho$ in our case;
the finding of $\sigma\sim\xi^2$ in Ref.\ \cite{HN} 
implies $\rho\sim\xi^2$ in our situation.\cite{Note1}

The spontaneous inter-valley coherence discussed above is very similar to
the inter-layer coherence in the double-layer $\nu=1$ QHE
\cite{QHF_review,Eisenstein_2004}. 
However, there are several important differences.
(i) The parameter $d/\ell_B$ ($d$ is the interlayer separation)
in the double-layer system is replaced by $a/\ell_B$
where $a$ is the lattice spacing.
For the current system $a/\ell_B\ll 1$, a regime which has not been achieved in the double layer system.
(ii)  The inter-valley coherent state we propose is spin singlet rather than spin polarized.
(From this point of view, the $\nu=1$ bilayer QH system is similar to
the $\nu=\pm 1$ QHE in graphene rather than $\nu=0$.)
Although the (pseudospin) supercurrent cannot be directly measured in the inter-valley coherent state in graphene, these two facts have advantage over double-layer QH systems to observe 
the KT physics.

We stress that our proposal is motivated by the apparent KT-divergent resistance and the highly resistive metallic state at $\nu=0$
observed in
Ref.\ [\onlinecite{Ong08PRL}]. 
The very basis of our proposal should be subjected to further experimental scrutiny, by changing temperature, magnetic field, doping, and mobility.

We are grateful to N.\ P.\ Ong and Y.\ Zhang for useful discussions.
KN thanks R. Saito and K. Sasaki for helpful arguments on electron-phonon coupling, 
and A.\ H.\ MacDonald and J. Sinova on the 
HF calculation.
He was supported by MEXT Grand-in-Aid No.20740167.
DHL was supported by DOE grant number DE-AC02-05CH11231.
SR thanks the Center for Condensed Matter Theory at University of California,
Berkeley for its support.

\end{document}